\documentclass[12pt]{article}     
\normalbaselines

\begin{document}
\newcommand{\pst}{\hspace*{1.5em}}
\newcommand{\be}{\begin{equation}}
\newcommand{\ee}{\end{equation}}
\newcommand{\ds}{\displaystyle}
\newcommand{\bdm}{\begin{displaymath}}
\newcommand{\edm}{\end{displaymath}}
\newcommand{\beq}{\begin{eqnarray}}
\newcommand{\eeq}{\end{eqnarray}}

\begin{center} {\Large \bf
\begin{tabular}{c}
OPTICAL PROPAGATOR OF QUANTUM SYSTEMS \\
IN THE PROBABILITY REPRESENTATION
\end{tabular}
 } \end{center}

\smallskip

\begin{center} {\bf Yakov A. Korennoy$^*$ and Vladimir I. Man'ko }\end{center}

\medskip

\begin{center}
{\it
P.~N.~Lebedev Physical Institute, Russian Academy of Sciences\\
Leninskii Prospect 53, Moscow 119991, Russia}

\smallskip

\end{center}

\begin{abstract}\noindent
The evolution equation for the propagator of the quantum system in
the optical probability representation (optical propagator) is
obtained. The relations between the optical and quantum
propagators for the Schr\"odinger equation and the optical
propagator of an arbitrary quadratic system are found explicitly.
\end{abstract}


\section{Introduction}
\pst Since the beginning of quantum mechanics there were attempts
at understanding  the notion of quantum states in terms of a
classical approach
\cite{debroglie1,debroglie2,debroglie3,bohm1,bohm2}. Within the
framework of this approach, a number of the so-called
quasiprobability distribution functions, such as the Wigner
function \cite{wig32}, the Husimi function \cite{husimi}, and the
Glauber--Sudarshan function \cite{glauber63,sud63}, later on
unified into a one-parametric family \cite{cahill}, were
suggested. But due to the Heisenberg principle, in contrast to the
classical probabilities, all these quasiprobabilities do not
describe distributions of measurable variables.

A formulation of quantum mechanics that is very similar to the
classical stochastic approach has been presented by Moyal
\cite{moyal}, but the evolution equation suggested by Moyal was an
equation for the quasiprobability distribution function (the
Wigner function) and not for the probability.

Recently in \cite{tom3,tom33}, it was suggested to consider
quantum dynamics as a classical stochastic process described
namely by the probability distribution -- so-called marginal
distribution function (MDF) (which was discussed in a general
context in \cite{cahill}) associated to the position coordinate
$X$, taking values in an ensemble of reference frames in the phase
space. Such a classical probability distribution is shown to
describe completely quantum states \cite{tom1,tom2,tom2_1}.
Detailed analysis of the tomographic probability picture of quantum mechanics was presented in \cite{Ibort}.

Within this approach, the notion of measuring a quantum state
provides the usual optical tomography approach
\cite{berber,vogrisk,raymer} and its extension called the
symplectic-tomography formalism \cite{tom1,tom2,tom2_1}. Both
allow one to link explicitly the MDF and the Wigner function or
the density matrix in other representations. In this way, starting
from the evolution equation for the density matrix, the evolution
equation of a Fokker--Plank type for the symplectic marginal
distribution function (symplectic tomogram) is obtained
\cite{tom3,tom33}. The quantum evolution equation and the energy
level equation have been found for the optical tomogram in an
explicit form in \cite{Kor13}. Such equations allows one to define
independently the marginal distribution function. Thus it may be
assumed as a starting point for an alternative but equivalent
formulation of quantum mechanics in what we call the probability
representation or the classical-like description of quantum
mechanics.

In \cite{manko1,manko1_1,manko1_2,OVManko1},
the symplectic propagator for
a wide class of Hamiltonians and its relation to the quantum
propagator for the density matrix was found, which establishes an
important bridge between the probability representation of quantum
mechanics and other formulations such as the path-integral
approach.

In the present work, we extend the optical probability description
of quantum mechanics and verify the formalism by applying it to a
nontrivial case of physical interest, which is the forced
parametric oscillator. From the viewpoint of general formalism, we
find new equations that connect the optical propagator of a
quantum system with its integrals of motion.

This paper is organized as follows.

In Sec. 2, we review the optical tomography of quantum states. In
Sec. 3, we find general relations for the optical propagator and
present the evolution equation for the optical propagator with the
initial conditions. In Sec. 4, we obtain the optical propagator of
an arbitrary quadratic system and check the general relations by
direct substitution of found propagators for simple quantum
systems.

\section{Optical Tomographic Representation of Quantum States}
\pst In this section, we give a short review of the tomographic
representation of quantum mechanics using the so-called optical
tomogram \cite{berber,vogrisk,raymer}. For simplicity, we consider
the case of one degree of freedom with dimensionless variables,
because the generalization of all our calculations and results to
the case of any arbitrary number of dimensional degrees of freedom
is obvious.

If $\hat\rho$ is the density matrix of the quantum state, the
optical tomogram is defined as follows:
\begin{equation}        \label{eq1}
w(X,\theta)= \mbox{Tr}\{\hat\rho\delta(X-\hat q\cos\theta-\hat p\sin\theta)\}
=\langle X,\theta\vert\hat\rho\vert X,\theta\rangle,
\end{equation}
where $\vert X,\theta\rangle$ is an eigenvector of the Hermitian
operator $\hat q\cos\theta+\hat p\sin\theta$ for the eigenvalue
$X$
\begin{equation}        \label{eq2}
\langle q\vert
X,\theta\rangle=\left({2\pi}\vert\sin\theta\vert\right)^{-1/2}
\exp\left(i\frac{Xq-\frac{q^2}{2}\cos\theta}{\sin\theta}\right).
\end{equation}

In terms of the Wigner function \cite{wig32}, the tomogram
$w(X,\theta)$ is expressed as
\begin{equation}        \label{eq4}
w(X,\theta)=\frac{1}{(2\pi)^2}\int W(q,p)e^{i\eta(X-q\cos\theta-
p\sin\theta)}dq\,dp\,d\eta.
\end{equation}
This relation can be reversed using the symmetry property
 of the optical tomogram
\be     \label{eq4_1} w(X,\theta,t)=w((-1)^k X,\theta+\pi
k,t),\qquad k=0,~\pm1,~\pm2, \cdots \ee After some algebra, we
arrive at
\begin{equation}        \label{eq5}
W(q,p)=\frac{1}{2\pi}\int\limits_{0}^{\pi}~\mbox{d}\theta
\int\limits_{-\infty}^{+\infty}\int\limits_{-\infty}^{+\infty}w(X,\theta)
|\eta |~e^{i\eta(X-q\cos\theta-p\sin\theta)} \mbox{d}\eta~\mbox{d}X.
\end{equation}
From (\ref{eq4}), using the relations between the Wigner function
and the density matrix $\rho(q,q')$ in the coordinate
representation
\begin{eqnarray}
&&W(q,p)=\int\rho(q+u/2,q-u/2)e^{-ipu}du, \label{eq6}\\
&&\rho(q,q')=\frac{1}{2\pi}\int W\left(\frac{q+q'}{2},p\right)
e^{ip(q-q')}dp,\label{eq7}
\end{eqnarray}
we can write the relations between the optical tomogram and the
density matrix in the coordinate representation as follows:
\begin{eqnarray}
&&w(X,\theta)=\frac{1}{2\pi}\int\rho\left(
q+\frac{u\sin\theta}{2},q-\frac{u\sin\theta}{2}
\right)e^{-iu(X-q\cos\theta)}du~dq,\label{eq8}\\
 &&\rho(q,q')=\frac{1}{2\pi}\int\limits_{0}^{\pi}~\mbox{d}\theta
\int\limits_{-\infty}^{+\infty} w(X,\theta)\vert\eta\vert
\exp\left\{i\eta\left(X-\frac{q+q'}{2}\cos\theta\right)\right\}
\delta(q-q'-\eta\sin\theta)~{\mbox d}\theta~d\eta~dX.\nonumber\\
&& \label{eq9}
 \end{eqnarray}
Thus, the tomogram $w(X,\theta)$ contains all information on the
quantum state.

The quadrature statistics can be obtained from the optical
tomogram
\begin{equation}        \label{eq10}
\langle X^n\rangle(\theta)=\int X^nw(X,\theta)~dX.
\end{equation}

\section{Quantum Evolution and Optical Propagator}

\pst The quantum evolution equation for the optical tomogram has
been found in explicit form in \cite{Kor13}. In the
one-dimensional case, for the Hamiltonian $\hat H=(\hat
p^2/2)+U(\hat q)$ this equation reads
\begin{eqnarray}
&&\frac{\partial}{\partial t}w(X,\theta,t)=
\left[\cos^2\theta\frac{\partial}{\partial\theta}
-\frac{1}{2}\sin2\theta\left\{1+X\frac{\partial}{\partial
X}\right\}
\right]w(X,\theta,t) \nonumber \\
&&+2\left[\mbox{Im}~U\left\{
\sin\theta\frac{\partial}{\partial\theta}
\left[\frac{\partial}{\partial X}\right]^{-1}
+X\cos\theta+i\frac{\sin\theta}{2} \frac{\partial}{\partial
X}\right\}\right] w(X,\theta,t).
        \label{eq11}
\end{eqnarray}
The evolution of the probability distribution $w(X,\theta)$ can be
represented as the following integral relationship:
\begin{equation}                             \label{eq12}
w(X,\theta,t)=\int\limits_0^\pi {\mbox d}\theta
\int\limits_{-\infty}^\infty {\mbox
d}X~\Pi(X,\theta,t;X',\theta',t') w(X',\theta',t'),\qquad t\geq
t',
\end{equation}
where $\Pi(X,\theta,t;X',\theta',t')$ is the Green function of Eq.
(\ref{eq11}). We call this function the optical
propagator. In the probability terminology, it can be interpreted
as the classical probability density of the system's transition
from the initial position $ X' $ in the ensemble of reference
frames of the classical phase space to the position $ X$.

Due to (\ref{eq4_1}), the optical propagator satisfies the
symmetry property \be     \label{eq_12_1}
\Pi\left(X,\theta,t;X',\theta',t'\right)=
\Pi\left((-1)^kX,\theta+\pi k,t;X',\theta',t'\right),\qquad t\geq
t'. \ee

Taking into account the physical meaning, the propagator satisfies
the nonlinear integral relationship in the case of $N$
consecutive moments of time $t_k$ ($k=\overline{1,~N}$) between
$t_1=t_{in}$ and $t_f=t_N$ 
\be \label{eq_12_2}
\Pi(X_N,\theta_N,t_f;X_1,\theta_1,t_{in})= \int\left\{\prod_{k=1}^{N-1}
\Pi(X_{k+1},\theta_{k+1},t_{k+1};X_k,\theta_k,t_k)\right\}\prod_{k=2}^{N-1}
\,dX_k\,d\theta_k.
\ee

If one takes in this relation $t_f-t_{in}=N\tau$,
$t_k=t_{in}+k\tau$, in the limit case $\tau\to0,$ $N\to\infty$,
the expression for the optical propagator can be obtained in terms
of the functional integral.

We establish the connection between the optical propagator and the
quantum propagator (Green function) for the density matrix
$\rho(q,q',t)$.

For the pure state with the wave function $\Psi(q,t)$, we have
\bdm \rho(q,q',t)=\Psi(q,t)\Psi^*(q',t). \edm Since the wave
function at the time moment $t$ is related to the wave function at
the time moment $t'$ by the Green function $G(q,q',t,t')$ of the
Schr\"odinger equation \bdm \Psi(q,t)=\int G(q,\tilde q,t,t')\Psi(\tilde
q,t')d\tilde q,\qquad t\geq t', \edm for the density matrix we can
write \bdm \rho(q,q',t)=\int K(q,q',t;\tilde q,\tilde q',t')
\rho(\tilde q,\tilde q',t')\,d\tilde q\,d\tilde q',\qquad t\geq
t', \edm where $K(q,q',t;\tilde q,\tilde q',t') =G(q,\tilde
q,t,t')G^*(q',\tilde q',t,t')$  is the quantum propagator for the
density matrix. Obviously, the quantum propagator satisfies the
initial condition \be \label{eq3_11} K(q,q',t';\tilde q,\tilde
q',t')=\delta(q-\tilde q)\delta(q'-\tilde q'). \ee Usually  the
quantum propagator is taken to be zero at $t<t'$: \be
\label{eq3_11_1} K(q,q',t;\tilde q,\tilde q',t')=0,\qquad t<t'. \ee

In view of the relation between the density matrix and the optical
tomogram (\ref{eq9}), we obtain
\begin{eqnarray}
&&K(q,q',t;\tilde q,\tilde q',t')= \frac{1}{(2\pi)^2}\int
\Pi\left(X,\theta,t; \frac{1}{2}(\tilde q+\tilde q')\cos\theta'+
p'\sin\theta',\theta',t'\right)
~e^{-ip'(\tilde q-\tilde q')} \nonumber \\
&&\times e^{i\eta[X-((q+q') \cos\theta)/2]}|\eta|
\delta(q-q'-\eta\sin\theta)\,d\eta\,dp'\,dX\,d\theta\,d\theta'.
\label{r13_1}
\end{eqnarray}
Thus, if we know the propagator for the optical probability
distribution $w(X,\theta)$, we can find the quantum propagator for
the density matrix of the quantum state.

Formula (\ref{r13_1}) can be reversed as follows:
\begin{eqnarray}
&&\Pi(X,\theta,t;X',\theta',t')=\frac{1}{(2\pi)^2} \int
K\left(q+\frac{k\sin\theta}{2}, q-\frac{k\sin\theta}{2},t; \tilde
q'+\eta\sin\theta', \tilde q',t'\right)\nonumber \\
&&\times\exp\left\{ i\eta\left(X'-\frac{2\tilde
q+\eta\sin\theta'}{2}
\cos\theta'\right)\right\}\exp\left\{-ik(X-q\cos\theta) \right\}
|\eta|\,dk\, d\eta\,dq\,d\tilde q'. \label{r13_2}
\end{eqnarray}
After substituting the initial condition (\ref{eq3_11}) for the
quantum propagator into expression (\ref{r13_2}), we can obtain
the initial condition for the optical propagator, namely, \be
\label{eq4_1_1} \Pi(X,\theta, t';X',\theta',t')=
\delta(X\cos(\theta-\theta')-X')\delta(\sin(\theta-\theta')). \ee
Obviously, this condition satisfies the symmetry property
(\ref{eq_12_1}).

In view of (\ref{eq12}), taking into account the initial condition
(\ref{eq4_1_1}), and defining  \bdm
\Pi(X,\theta,t;X',\theta',t')=0,\quad \mbox{at}\quad t<t', \edm
from (\ref{eq11}) we can obtain that the optical propagator
satisfies the equation
\begin{eqnarray}
&&\frac{\partial}{\partial t}\Pi(X,\theta,t;X',\theta',t')-
\left[\cos^2\theta\frac{\partial}{\partial\theta}
-\frac{1}{2}\sin2\theta\left\{1+X\frac{\partial}{\partial
X}\right\} \right]\Pi(X,\theta,t;X',\theta',t') \nonumber\\
&&-2\left[\mbox{Im}~U\left\{
\sin\theta\frac{\partial}{\partial\theta}
\left[\frac{\partial}{\partial X}\right]^{-1}
+X\cos\theta+i\frac{\sin\theta}{2} \frac{\partial}{\partial
X}\right\}\right]
\Pi(X,\theta,t;X',\theta',t') \nonumber \\
&&=\delta(X\cos(\theta-\theta')-X')\delta(\sin(\theta-\theta'))\delta(t-t').
        \label{eq13}
\end{eqnarray}

It is worth noting that relations (\ref{r13_1})--(\ref{r13_2}) are
the extensions of known results \cite{DSF-57_97} for the
symplectic tomogram \be \label{eq13_11}
M(X,\mu,\nu,t)=\frac{1}{(2\pi)^2}\int W(q,p,t)e^{ik(X-\mu q-\nu
p)}\,dk\,dq\,dp, \ee in the case where the symplectic propagator
is defined as \be \label{r68_1} M(X,\mu,\nu,t)=\int
\Pi(X,\mu,\nu,t;X',\mu',\nu';t') M(X',\mu',\nu',t')\,dX'\,
d\mu'\,d\nu'. \ee The relation between the symplectic tomogram and
the quantum propagator reads
\begin{eqnarray}
K(q,q',t;\tilde q,\tilde q',t')&=&\frac{1}{(2\pi)^{2}}
\int\frac{1}{|\nu'_\sigma|}\exp\left\{
iY-i\mu\frac{q+q'}{2}
-iY'\frac{\tilde q-\tilde q'}{\nu'}
+i\mu'\frac{\tilde q^2-\tilde q'^2}{2\nu'}
\right\}\nonumber \\
{}&\times& \Pi(Y,\mu,q-q',t;Y',\mu',\nu';t')\,d\mu\, d\mu'\,
dY\,dY'\,d\nu'. \label{eq17}
\end{eqnarray}
For the inverse relation we have
\begin{eqnarray}
\Pi(X,\mu,\nu,t;X',\mu',\nu';t')&=&
\frac{1}{(2\pi)^2|\nu|}\int
\exp\left\{-i\frac{q-q'}{\nu}
\left(X-\mu\frac{q+q'}{2}\right)
+iX'-i\mu'\frac{2\tilde q-\nu'}{2}
\right\}\nonumber \\
{}&\times&K(q,q',t;\tilde q,\tilde q-\nu',t') \,dq\,dq'\,d\tilde
q. \label{eq17_1}
\end{eqnarray}
Using the homogeneity property of the symplectic tomogram \be
\label{eq18_1} M(\lambda X,\lambda\mu,\lambda\nu,t)=
\vert\lambda\vert^{-1}M(X,\mu,\nu,t), \ee we can rewrite this
relation as follows:
\begin{eqnarray}
&&\Pi(X,\mu,\nu,t;X',\mu',\nu';t')= \frac{1}{(2\pi)^2|\nu|}\int
\exp\left\{i\frac{\nu_1}{\nu} \left(X'-\mu'\frac{2\tilde
q-\nu_1}{2} \right)\right\}\nonumber\\
{}&&\times\exp\left\{-i\frac{q-q'}{\nu}
\left(X-\mu\frac{q+q'}{2}\right)\right\} \vert\nu_1\vert
\delta(\nu-\nu')K(q,q',t;\tilde q,\tilde
q-\nu_1,t')\,dq\,dq'\,d\tilde q\,d\nu_1. \label{eq19_1}
\end{eqnarray}
Also the property (\ref{eq18_1}) enables us to find the relation
between the symplectic and optical propagators: \be \label{eq20_1}
\Pi_{\mbox{\footnotesize opt}}(X,\theta,t;X',\theta',t')=\int
\Pi_{\mbox{\footnotesize s}}(X,\cos\theta,\sin\theta,t; r
X',r\cos\theta', r\sin\theta';t')\vert r\vert\,dr. \ee

\section{Optical Propagator for Quadratic Systems}

\pst As an example, we consider the system with quadratic Hermitian
Hamiltonian \bdm \hat H=\frac{1}{2}(\hat{\bf Q}B\hat{\bf Q})+{\bf
C}\hat{\bf Q}, \edm where  $\hat{\bf Q}=(\hat p,\hat q)$ is a
vector operator, $B$ is a symmetric 2$\times$2 matrix, and ${\bf
C}$ is a real 2-vector dependent on time. As known
\cite{Laser41,Laser42}, the system has linear integrals of motion:
\be                             \label{r47_4} \hat{\bf
I}(t)=\Lambda(t)\hat{\bf Q}+{\bf\Delta}(t), \ee where the real
symplectic 2$\times$2 matrix $\Lambda(t)$ and the real vector
${\bf\Delta}(t)$ satisfy the equations \bdm \dot\Lambda=i\Lambda
B\sigma_y,\qquad\dot{\bf\Delta}=i{\bf\Delta}\sigma_y{\bf C}, \edm
with the initial conditions $\Lambda(0)=1$ and ${\bf\Delta}(0)=0.$
The integrals of motion are defined by the following equation: \be
\label{r32_1}
\partial_t \hat I(t)+i[\hat H,\hat I(t)]=0.
\ee

For the quadratic systems under consideration, any integrals of
motion can be expressed as functions of two operators: $\hat
A(t)=\hat U\hat a\hat U^{-1}$ and
$\hat A^{\dag}(t)=\hat U\hat a^{\dag}\hat U^{-1},$ where $\hat U$ is the evolution operator and
\begin{equation}
\hat A(t)=\frac{i}{\sqrt2}(\varepsilon(t)\hat
p-\dot\varepsilon(t)\hat q) +\beta(t), \qquad \hat
A^{\dag}(t)=-\frac{i}{\sqrt2}(\varepsilon^*(t)\hat
p-\dot\varepsilon^*(t)\hat q) +\beta^*(t), \label{r48_1}
\end{equation}
with $\varepsilon$ satisfying the equations \be \label{r48_2}
\ddot\varepsilon+\omega^2(t)\varepsilon=0,\qquad
\dot\varepsilon\varepsilon^*-\dot\varepsilon^*\varepsilon=2i, \ee
with the initial conditions $\varepsilon(0)=1$ and
$\dot\varepsilon(0)=i.$ The function $\beta(t)$ is defined from
(\ref{r32_1}) as  \be \label{r48_3}
\beta(t)=-\frac{i}{\sqrt2}\int\limits_0^t{\mbox
d}t'\varepsilon(t')f(t'), \ee  and the integrals of motion
(\ref{r47_4}) are expressed from $\hat A$ and $\hat A^{\dag}$ as
follows: \be \label{r48_4} \hat{\bf I}(t)=\left(\hat I_p\atop \hat
I_q\right)= \left(\frac{\hat A-\hat A^{\dag}}{i\sqrt2}\atop
\frac{\hat A+\hat A^{\dag}}{\sqrt2}\right),\qquad \hat{\bf
I}(t=0)=\left(\hat p\atop \hat q\right). \ee The matrix $\Lambda$
and the vector ${\bf\Delta}$ read \bdm \Lambda=\frac{1}{2}
\left({\varepsilon+\varepsilon^*\atop
i(\varepsilon-\varepsilon^*)}\quad
{-(\dot\varepsilon+\dot\varepsilon^*)\atop
-i(\dot\varepsilon-\dot\varepsilon^*)}\right),\qquad
{\bf\Delta}=\frac{1}{\sqrt2}\left(i(\beta-\beta^*)\atop
\beta+\beta^*\right). \edm The knowledge of integrals of motion
(\ref {r48_4}) allows one to find the Green function (or quantum
propagator) for the Schr\"odinger equation of the system (see
\cite{Laser41}).

Now we expand the method of integrals of motion in order to find
the optical propagator. Taking into account the initial condition
(\ref{eq4_1_1}), we can write the system of equations (we put
$t'=0$)
\begin{equation} \tilde
q\Pi(X,\theta,X',\theta',t=0)=\tilde
q'\Pi(X,\theta,X'\theta',t=0),\qquad \tilde
p\Pi(X,\theta,X',\theta',t=0)=-\tilde
p'\Pi(X,\theta,X'\theta',t=0), \label{eq191}
\end{equation}
where  $\tilde p$ and $\tilde q$ are the quadrature operators in
the optical tomographic representation \cite{Kor13},
\begin{eqnarray}
\tilde q&=&\sin\theta\left[\frac{\partial}{\partial X}\right]^{-1}
\frac{\partial}{\partial\theta}+X\cos\theta+\frac{i}{2}
\sin\theta\frac{\partial}{\partial X},
\nonumber \\[-2mm]
&&\label{eq_p38}\\[-2mm]
\tilde p&=&-\cos\theta\left[\frac{\partial}{\partial
X}\right]^{-1}
\frac{\partial}{\partial\theta}+X\sin\theta-\frac{i}{2}
\cos\theta\frac{\partial}{\partial X}. \nonumber
\end{eqnarray}
Here, the primes mean the action to the primed variables of the
propagators.

During the time evolution (or formal action of the evolution
operator to nonprimed variables), the system (\ref{eq191}) evolves
to the following one:
\begin{eqnarray}
\tilde{\bf I}_q(t)\Pi(X,\theta;X',\theta',t)&=&
\tilde q'\Pi(X,\theta;X',\theta',t), \label{r49_1} \\
\tilde{\bf I}_p(t)\Pi(X,\theta;X',\theta',t)&=& -\tilde
p'\Pi(X,\theta;X',\theta',t), \label{r49_2}
\end{eqnarray}
where the operators $\tilde{\bf I}_p$ and $\tilde{\bf I}_q$ in the
optical tomographic representation correspond to the operators
(\ref{r48_4}):
\begin{eqnarray}
\tilde{\bf I}_q&=&\frac{1}{4}[\cos\theta(\varepsilon-\varepsilon^*)+
\sin\theta(\dot\varepsilon-\dot\varepsilon^*)]\frac{\partial}{\partial X}+
\frac{i}{2}\left[\frac{\partial}{\partial X}\right]^{-1}
\left[(\dot\varepsilon-\dot\varepsilon^*)\left(-\sin\theta
\frac{\partial}{\partial\theta}-\cos\theta\left(
1+X\frac{\partial}{\partial X}\right)\right)\right.\nonumber \\
{}&&-\left.(\varepsilon-\varepsilon^*)\left(\cos\theta\frac{\partial}{\partial\theta}-
\sin\theta\left(1+X\frac{\partial}{\partial X}\right)\right)\right]+
\frac{\beta+\beta^*}{\sqrt2}, \nonumber \\
\tilde{\bf I}_p&=&-\frac{i}{4}[\cos\theta(\varepsilon+\varepsilon^*)+
\sin\theta(\dot\varepsilon+\dot\varepsilon^*)]\frac{\partial}{\partial X}+
\frac{1}{2}\left[\frac{\partial}{\partial X}\right]^{-1}
\left[(\dot\varepsilon+\dot\varepsilon^*)\left(-\sin\theta
\frac{\partial}{\partial\theta}-\cos\theta\left(
1+X\frac{\partial}{\partial X}\right)\right)\right.\nonumber \\
{}&&-\left.(\varepsilon+\varepsilon^*)\left(\cos\theta\frac{\partial}{\partial\theta}-
\sin\theta\left(1+X\frac{\partial}{\partial X}\right)\right)\right]+
\frac{\beta-\beta^*}{\sqrt2}. \nonumber
\end{eqnarray}

One can solve the system of equations  (\ref{r49_1}),
(\ref{r49_2}) by the following method.

First, from (\ref{r49_1}) the dependence on nonprimed variables is
completely obtained, considering the primed variables as
parameters, then from (\ref{r49_2}) the dependence on primed
variables is found. The time-dependent factor is obtained by the
substitution of the propagator to the dynamic equation
(\ref{eq13}) with the initial condition (\ref{eq4_1_1}).

In our case, Eq.  (\ref{eq13}) has the form
\begin{eqnarray}
&&\dot\Pi-(\cos^2\theta+\omega^2(t)\sin^2\theta)\partial_\theta\Pi+
(1-\omega^2(t))\sin\theta\cos\theta(1+X\partial_X)\Pi
+f(t)\sin\theta\partial_X\Pi\nonumber\\
&&=\delta(t)\delta(X\cos(\theta-\theta')-X')\delta(\sin(\theta-\theta')).
\label{eq1911}
\end{eqnarray}

The described procedure enables us to find the optical propagator
for our quantum system, but we can also find it from the results
of \cite{DSF-57_97} where the symplectic propagator of the
parametric-driven quadratic system was found, \be \label{eq1911_1}
\Pi(X,\mu,\nu;X',\mu',\nu')=\delta(X-X'+{\cal
N}\Lambda^{-1}\Delta) \delta({\cal N}'-{\cal N}\Lambda^{-1})
\Theta(t), \ee where ${\cal N}$ and ${\cal N}'$ are vectors ${\cal
N}=(\nu,\mu)$ and ${\cal N}'=(\nu',\mu')$, and $\Theta(t)$ is a
Heaviside step function \bdm
\Theta(t)=\left\{{1,~~\mbox{at}~~t\geq 0,}\atop {0,~~\mbox{at}~~t
< 0.}\right. \edm With the help of (\ref{eq20_1}), we can write
\be \label{r49_5} \Pi(X,\theta;X',\theta',t)=\delta\{{\cal
N}'_2(X+{\cal N}\Lambda^{-1}{\bf\Delta})- X'{\cal
N}\Lambda_{II}^{-1}\}~ \delta\{{\cal N}'_1-{\cal N}_2{\cal
N}\Lambda_I^{-1}/ {\cal N}\Lambda_{II}^{-1}\}\Theta(t), \ee where
the vectors  ${\cal N}=(\sin\theta,\cos\theta)$ and ${\cal
N}'=(\sin\theta',\cos\theta')$, now $\Lambda_I^{-1}$ and
$\Lambda_{II}^{-1}$ are the first and second columns of the matrix
$\Lambda^{-1},$ inverse to $\Lambda,$ and the indices 1 and 2 are
related to the first and second coordinates of the vectors. In
view of the time-dependent functions $\varepsilon(t)$ and
$\beta(t)$, the propagator $\Pi(x,\theta;x',\theta',t)$ is
expressed as follows:
\begin{eqnarray}
&&\Pi(X,\theta;X',\theta',t)=\delta[X\cos\theta'
-X'(\sin\theta(\dot\varepsilon^*+\dot\varepsilon)
+\cos\theta(\varepsilon^*-\varepsilon))/2\nonumber\\
{}&&+
\cos\theta'(\sin\theta(\dot\varepsilon^*\beta+\dot\varepsilon\dot\beta)+
\cos\theta(\varepsilon^*\beta+\varepsilon\beta^*))/\sqrt2]\nonumber\\
&&\times \delta\left[i\cos\theta'\frac{
\sin\theta(\dot\varepsilon^*-\dot\varepsilon)
+\cos\theta(\varepsilon^*-\varepsilon)}
{\sin\theta(\dot\varepsilon^*+\dot\varepsilon)
+\cos\theta(\varepsilon^*+\varepsilon)}-\sin\theta'
\right]\Theta(t).\nonumber
\end{eqnarray}
Thus if the integrals of motion are known, e.g., the matrix
$\Lambda(t)$ and the vector ${\bf\Delta}(t)$ are known, then
according to formula (\ref{r49_5}), the optical propagator is
known. In the particular case for a free motion  (with the
Hamiltonian $\hat H=\hat p^2/2$), we have \cite{Laser41,Laser42}
\bdm p_o(t)=p,\qquad q_o(t)=q-pt,\qquad {\bf\Delta}(t)=0,\qquad
\Lambda(t)=\left(~1~~0\atop-t~~1\right) \edm and the propagator of
free motion reads \be                             \label{r50_3}
\Pi_{\mbox{\footnotesize
f}}(X,\theta;X',\theta',t)=\delta(X\cos\theta'-X'\cos\theta)
\delta(\cos\theta'(t+\tan\theta)-\sin\theta')\Theta(t). \ee For
the harmonic oscillator with the Hamiltonian $\hat H=\hat
p^2/2+\hat q^2/2$ \bdm \Lambda(t)=\left(\cos t~~~\sin t\atop -\sin
t~~\cos t\right), \edm from (\ref{r49_5}) we obtain \be
\label{r50_5} \Pi_{\mbox{\footnotesize
os}}(X,\theta;X',\theta',t)=\delta\left(X\cos(\theta-\theta'+t)
-X'\right)\delta(\sin(\theta-\theta'+t))\Theta(t). \ee The
evolution equation, in this case, is very simple, \be
\label{r50_6}
(\partial_t-\partial_\theta)\Pi(X,\theta;X',\theta',t)=
\delta(t)\delta(X\cos(\theta-\theta')-X')\delta(\sin(\theta-\theta')),
\ee or for the optical distribution \bdm
(\partial_t-\partial_\theta)w_{\mbox{\footnotesize
os}}(X,\theta,t)=0. \edm

It is easily to check that (\ref{r50_3}) and (\ref{r50_5}) satisfy
the nonlinear expression (\ref{eq_12_2}), and substitution of
(\ref{r50_3}) and (\ref{r50_5}) into (\ref{r13_1}) provides known
results for quantum propagators  (see Appendix)
\begin{eqnarray}
K_{\mbox{\footnotesize f}}(q,q',\tilde q,\tilde q',t)&=&
\frac{1}{2\pi t}\exp\left[\frac{i}{2t}(q-\tilde q)^2
-\frac{i}{2t}(q'-\tilde q')^2\right]\Theta(t);\label{K_f} \\
K_{\mbox{\footnotesize os}}(q,q',\tilde q,\tilde q',t)&=&
\frac{1}{2\pi\vert\sin t\vert}\exp\left[\frac{i}{2}(q^2-q'^2
+\tilde q^2-\tilde q'^2)\cot t
-\frac{i}{\sin t}(q\tilde q - q'\tilde q')\right]\Theta(t).\label{K_os}
\end{eqnarray}
Substituting these expressions to (\ref{r13_2}) gives us
(\ref{r50_3}) and (\ref{r50_5}), respectively.
 Thus, we have
checked the expressions obtained in the present paper.

Note that during the procedure of dimension restoration, the
Planck's constant $\hbar$ appears neither in the dynamic equation
for the propagator and the marginal distribution of quadratic
system nor in the general expression for the propagator; in this
sense, the dynamic equation and the propagator are classical (if,
of course, $ \omega $ and $f $ do not depend on $ \hbar $).
However, the initial condition for the dynamic equation of
marginal distribution can contain the Planck's constant, i.e., to
be quantum.

\section{Conclusions}
\pst To summarize, we point out the main results of this study.

We obtained the relations between the propagators for the optical
tomogram and for the density matrix of quantum systems. We
presented the evolution equation for the optical propagator along
with the initial conditions. We showed the correspondence between
the optical and symplectic propagators. As an example, we derived
the optical propagators of arbitrary quadratic systems and checked
our general expressions by direct substitution. The importance of
the expressions written here for optical tomographic
representation is connected with the fact that namely these
tomograms are measured in the experiments on photon states where
their characteristics are studied. The propagators and the
evolution equation found in our work provide the possibility of
monitoring the system quantum states in the process of time
evolution.

\section{Appendix}

Substituting (\ref{r50_5}) to (\ref{r13_1}) we can write
\bdm
\frac{1}{(2\pi)^2}\int \delta\left( X\cos(\theta-\theta'+t)-
\frac{1}{2}(\tilde q+\tilde q')\cos\theta'-p'\sin\theta'\right)
\delta(\sin(\theta-\theta'+t))
\exp(-ip'(\tilde q-\tilde q'))
\edm
\bdm
\times\exp\left(i\eta(X-\frac{1}{2}(q+q')\cos\theta)\right)
\vert\eta\vert\delta(q-q'-\eta\sin\theta)
\mbox{d}\eta~\mbox{d}p'~\mbox{d}X~\mbox{d}\theta~\mbox{d}\theta'=
\edm
\bdm
=\frac{1}{(2\pi)^2}\int \delta\left( X(-1)^k-
\frac{1}{2}(\tilde q+\tilde q')(-1)^k\cos\theta'-p'(-1)^k\sin\theta'\right)
\exp(-ip'(\tilde q-\tilde q'))
\edm
\bdm
\times\exp\left(i\eta(X-\frac{1}{2}(q+q')\cos\theta)\right)
\vert\eta\vert\delta(q-q'-\eta\sin\theta)
\mbox{d}\eta~\mbox{d}p'~\mbox{d}X~\mbox{d}\theta=
\edm
\bdm
=\frac{1}{(2\pi)^2}\int 
\exp(-ip'(\tilde q-\tilde q'))
\vert\eta\vert\delta(q-q'-\eta\sin\theta)
\edm
\bdm
\times\exp\left\{i\eta\left(\frac{1}{2}(q+q')\cos(t+\theta)+p'\sin(t+\theta)
-\frac{1}{2}(q+q')\cos\theta\right)\right\}
\mbox{d}\eta~\mbox{d}p'~\mbox{d}\theta
\edm
After change of variables $\eta\cos\theta=y$, $\eta\sin\theta=z$
and calculations we arrive at (\ref{K_os}).

\end{document}